\documentclass[preprint,12pt]{elsarticle}
\usepackage[utf8]{inputenc}
\usepackage{amssymb}
\usepackage{geometry}
\usepackage{multirow}
\usepackage[section]{placeins}
\usepackage{amsmath}
\usepackage{lineno}
\journal{Information Sciences}
\usepackage[dvipsnames]{xcolor}
\usepackage{natbib}
\setcitestyle{authoryear,open={(},close={)}}
\usepackage[colorlinks=true]{hyperref}

\newcommand{\eref}[1]{Eq.~(\ref{#1})}
\newcommand{\ee}{\mathrm{e}}

\newcommand{\WinRatio}{\mathsf{WinRatio}}
\newcommand{\PageRank}{\mathsf{PageRank}}
\newcommand{\BiPageRank}{\mathsf{BiPageRank}}
\begin{document}
\begin{frontmatter}
\title{Limits of PageRank-based ranking methods in sports data}

\author[1]{Yuhao Zhou}
\author[1]{Ruijie Wang}
\author[1]{Yi-Cheng Zhang}
\author[1,2]{An Zeng\corref{co1}}
\cortext[co1]{Corresponding author}
\ead{anzeng@bnu.edu.cn}
\author[3,4,1]{Matúš Medo\corref{co2}}
\cortext[co2]{Corresponding author}
\ead{matus.medo@unifr.ch}

\address[1]{Department of Physics, University of Fribourg, Fribourg 1700, Switzerland}
\address[2]{School of Systems Science, Beijing Normal University, Beijing 100875, PR China}
\address[3]{Institute of Fundamental and Frontier Sciences, University of Electronic Science and Technology of China, Chengdu 610054, PR China}
\address[4]{Department of Radiation Oncology, Inselspital, University Hospital of Bern, and University of Bern, Bern 3010, Switzerland}

\begin{abstract}
While PageRank has been extensively used to rank sport tournament participants (teams or individuals), its superiority over simpler ranking methods has been never clearly demonstrated. We use sports results from 18 major leagues to calibrate a state-of-art model for synthetic sports results. Model data are then used to assess the ranking performance of PageRank in a controlled setting. We find that PageRank outperforms the benchmark ranking by the number of wins only when a small fraction of all games have been played. Increased randomness in the data, such as intrinsic randomness of outcomes or advantage of home teams, further reduces the range of PageRank's superiority. We propose a new PageRank variant which outperforms PageRank in all evaluated settings, yet shares its sensitivity to increased randomness in the data. Our main findings are confirmed by evaluating the ranking algorithms on real data. Our work demonstrates the danger of using novel metrics and algorithms without considering their limits of applicability.
\end{abstract}

\end{frontmatter}


\section{Introduction}
Humans are born to compete: We thrive by measuring ourselves against the others~\citep{frederick2003competition,le2013imaging}. Sport in particular provides ample opportunities for competition with high significance for the economy~\citep{tranter1998sport,hone2006measuring,rosner2011business} and the society~\citep{mcpherson1989social,giulianotti2015sport}. Once a sport competition is over, it remains to decide who has won. What is a simple question for a single match between two participants (individuals or teams) becomes highly non-trivial for a structured tournament with multiple games between several participants. The design of sport tournaments is therefore of crucial importance. Effective tournament design helps the participants to perform well during the tournament, produces match-ups that are interesting to the fans and, crucially, allows to identify the best-performing participant with high probability~\citep{szymanski2003economic} (see~\citep{dechenaux2015survey} for a survey of results on sport tournaments and beyond).

We focus here on sports leagues, such as a soccer league, for example, where each participating team plays against every other team once or several times. In such a league, points are traditionally assigned to teams for every win or tie that they achieve. The teams are then ranked by their point totals (from the highest to the lowest; additional criteria can be used to break ties). This benchmark ranking method is so simple that it inevitably leads to the question: Cannot we rank the teams better? Of particular appeal is the idea to consider the strength of the opposing team. In particular: Can we improve the ranking if a win against a strong team counts more than a win against a weak team? A closely related idea has been previously formalized by the PageRank algorithm which has been originally designed to rank web pages and later used in a broad range of systems~\citep{gleich2015pagerank}. Unsurprisingly, PageRank and its modifications have been also applied to sports results mapped onto a directed network where team $i$ losing against team $j$ is represented by an edge from $i$ to $j$~\citep{radicchi2011best,junior2012time,spanias2013tennis,lazova2015pagerank}.

While PageRank is frequently used on sports results data, its superiority over simple point-based schemes has not been established yet. That is not surprising as real sports results lack robust ground truth (a correct team ranking) against which different rankings could be compared. We fill this gap by providing an extensive evaluation of three different rankings algorithms (ranking by points, PageRank, and a new PageRank variant) on synthetic data. We first calibrate our data-generating model on real sports results from numerous leagues in four major sports (baseball, basketball, ice hockey, and soccer) and find that each sport has a characteristic range of model parameters. We then compare the performance of the ranking algorithms on synthetic data generated for various values of the model parameters, thus identifying the parameter ranges that are favorable for each of the algorithms. We establish in this way, for the first time, the conditions that have to be met for PageRank to perform better than a point-based ranking. We find that PageRank requires results with little randomness which is, in fact, not typical for real sports results. While the newly proposed PageRank variant outperforms PageRank in all evaluated settings, the point-based ranking is the best algorithm in most relevant settings. We conclude this work by reproducing our key results on real data, thus showing that our findings are not model-dependent.

This paper is structured as follows. Section 2 reviews modeling sports results data and ranking algorithms for sports results. Section 3 introduces the methods used in our work---the algorithm to create synthetic data, ranking algorithms, evaluation metrics, and real datasets. Section 4 describes model calibration on the real datasets. Sections 5 and 6 present a comparison of ranking algorithms on synthetic and real datasets, respectively. Finally, Section 7 summarizes the results, discusses their limitations, and proposes new research directions.

\section{Related work}

\subsection{Modeling sports results}
Results in a sport where players or teams play against each other can be seen as outcomes of paired comparisons between the participants. The Bradley-Terry model~\citep{bradley1952rank} is one of the first efforts to model outcomes of such paired comparisons (the authors themselves do not mention applying their model on sport specifically). The model is based on assigning a non-negative winning propensity, $\pi$, to each participant and postulating that the probability of $i$ winning against $j$ in the form $\pi_i/(\pi_i+\pi_j)$. A model generalization to include ties (which is important for sports such as soccer) has been introduced in~\citep{rao1967ties}. The Bradley-Terry model was used in a broad range of problems. In~\citep{aoki2017luck}, for example, the authors used the model to show that sports data have a high degree of randomness. In~\citep{deng2012universal}, the winning propensities $\pi$ of tennis players were shown to be directly proportional to players' ranks in official tennis rankings. The rank of individuals instead of a continuous-valued winning propensity was used also in~\citep{chetrite2017number}.

In~\citep{ben2013randomness}, the authors propose a simpler model based on a fixed upset probability parameter that directly specifies the probability that a weaker team beats a stronger team. The model's simplicity makes it possible to derive analytical results for the probability that the weakest team wins an elimination tournament, for example. \citet{o2008probability} goes in the opposite direction by proposing a model for tennis match outcomes based on the detailed structure of the game. See \citep{bradley1976science,david1988method,cattelan2012models} for detailed reviews of models for paired comparison data.

\subsection{Ranking algorithms for sports results data}
The most elementary method to aggregate results of multiple sports games is to compute each team's winning percentage, $n_w / (n_w + n_l)$, where $n_w$ and $n_l$ are the team's numbers of wins and losses, respectively. In this way, each team is assigned a quantity in the range $[0,1]$. Taking into account a starting uniform prior in the same range, a modified estimate has the form $(1+n_w) / (n_w+n_l+2)$. In~\citep{colley2002colley}, this estimate was used as a basis for a ranking scheme which is particularly useful when many teams have not played against each other (early in a season or in a more complicated setup where teams are divided in multiple divisions or conferences). The information provided by respective wins and losses is limited in low-scoring sports such as soccer where a single lucky shot can greatly influence the outcome of a match. In~\citep{brechot2020dealing}, the authors propose to limit this randomness by estimating the number of expected goals.

Indirect comparisons (comparing teams A and C based on team A beating team B and team B beating team C) have been considered in~\citep{redmond2003natural}. Indirect wins and indirect losses have been quantified in~\citep{park2005network} where they are ultimately combined in a score which is, in fact, a generalization of the well-known Katz centrality metric. A method based on representing the available results with an incomplete pairwise comparison matrix has been proposed in~\citep{bozoki2016application} where, however, the eventually optimized cost function has to be chosen from several distinct possibilities. In~\citep{spanias2013tennis}, a sort-based ranking has been proposed which also alleviates the problem of missing pair comparisons.

A popular line of research considers the use of eigenvector-based methods for sports rankings~\citep{keener1993perron}. In particular PageRank, a seminal ranking algorithm/centrality metric for nodes in a directed network~\citep{brin1998anatomy}, has been widely applied to sports data such as tennis~\citep{radicchi2011best} or cricket~\citep{mukherjee2012identifying}. PageRank-like algorithms seem well-suited for a sports ranking as they value a win against a strong opponent more than a win against a weak opponent. How to transform input sports results in a directed network on which PageRank is computed is open. The simplest approach is to represent a win of $i$ over $j$ with a directed link from $j$ to $i$. In~\citep{govan2008generalizing}, the authors proposed to assign link weights based on the score difference in the corresponding games. A comparison of various edge-weighting methods for the soccer World Cup data is presented in~\citep{lazova2015pagerank}. Time-aware PageRank variants~\citep{junior2012time,motegi2012network} take the time of each game into account to capture the player/team capability that varies in time.

Validation of the various ranking methods described above is often limited as it typically relies on official rankings that are directly influenced by the same results data that are used by the evaluated algorithm (see \citep{mukherjee2012identifying,junior2012time,lazova2015pagerank}, for example). However, the best agreement with an official ranking is achieved by a ranking method which is identical to that used to produce the official ranking.

\section{Materials and methods}

\subsection{Sports results model}
\label{sec:model}
We assume a competition setting where $N$ teams play against each other once or multiple times. In the model, we assume that the outcome of a match between teams $i$ (home team) and $j$ (away team) is stochastic. The probability that the home team wins is assumed in the form of the logistic function
\begin{equation}
\label{prob_win}
P(i,j)=1/\bigg[1+\ee^{-(f_{i}-f_{j}+H)/\delta}\bigg]^{-1}
\end{equation}
where $f_i$ and $f_j$ are the intrinsic fitness values of the two competing teams, $H$ is an additive term which represents the typical home team advantage and $\delta$ is a fitness ``weighting'' parameter which helps to translate a difference in team fitness in the winning probability of the home team. The model assumes that there are only two possible outcomes: team $i$ wins or team $j$ wins. Generalizing the model to involve the possibility of a draw or formulating a probability distribution for the score difference is beyond the scope of this article. We assume for simplicity that team fitness remains the same throughout the whole competition; allowing for fitness variations is yet another interesting direction for future research. Home advantage has been documented for a wide variety of sports~\citep{nevill2005extent,ribeiro2016advantage}. While it may seem as an auxiliary issue that has been ignored in \citep{deng2012universal}, for example, we find $H$ to be significantly positive in a vast majority of the sports results sets that we analyze in this paper. We also find that home advantage strongly affects the ranking ability of respective algorithms.

It is helpful to study closer the implications \eref{prob_win} before proceeding. When $(f_j-f_i)/\delta\gg1$ (that is, the away team is much stronger than the home team), we get $P_{ij}\ll1$ as expected. If $\delta$ increases, the same fitness difference affects the winning probability $P(i,j)$ less and the match outcomes thus become more random (in the limit $\delta\to\infty$, $P(i,j) = 1/2$ for any $f_i$, $f_j$, and $H$). We thus refer $\delta$ as the sport randomness indicator: When $\delta$ is large in comparison with fitness differences among the teams, $P(i,j)\approx 1/2$ for all $i$ and $j$. While home team advantage increases with $H$, outcomes are random in the large $\delta$ limit and no home team advantage ensues. The effective strength of home team advantage is thus determined by the ratio $H/\delta$.

\subsection{Algorithm to generate synthetic sports results}
The algorithm has the following parameters: number of teams $N$, fitness sensitivity $\delta$, home advantage $H$, and the fraction of games that have been played $P$ ($P=0$ and $P=1$ corresponds to no games played and all $N(N-1)/2$ games played, respectively). It is also possible to consider $P>1$: that would correspond to a league where the teams play more than once against each other.

Synthetic data are then created in three main steps:
\begin{enumerate}
\item Fitness of team $i$ is set to $f_i = (i - 0.5) / N$ where $i=1,\dots,N$. In this way, the fitness values range from $0.5/N$ to $1-0.5/N$ and they are regularly distributed in the range $[0,1]$ (we investigate other fitness distributions later).
\item Each team is assigned to play $P(N-1)$ games against opponents chosen at random without any two teams playing against each other more than once (in practice, we used the random\_degree\_sequence\_graph function from NetworkX~\citep{SciPyProceedings_11}). If $P(N-1)$ is not an integer, teams are assigned to play either $\lfloor P(N-1)\rfloor$ or $\lceil P(N-1)\rceil$ games in such a way that the total number of played games is $PN(N-1)/2$. The home team is chosen at random for each game.
\item Determine the outcome of each game by \eref{prob_win}.
\end{enumerate}
By varying the model parameters, we can create synthetic results corresponding to a broad range of sports. Note that the algorithm can be easily modified to encompass more complicated settings such as the regular season followed by playoffs, for example.

\subsection{Ranking algorithms}
In sports leagues, teams are typically ranked by the number of points that they obtain (such as two points for a win, one point for a draw, zero points for a loss). Since we focus here on sports where draws are not possible, ranking the teams by the number of points is the same as ranking them by the ratio of wins. $\WinRatio$ is thus our benchmark method.

To apply the PageRank algorithm, we create a directed network of participating teams where all games are represented with directed links. In particular, when player/team $i$ wins against player/team $j$, a directed link from $j$ to $i$ is formed along which ``sports prestige'' flows: A win against a highly-valued team contributes highly to the winner's own evaluation. The process can be mathematically represented by the formula~\citep{radicchi2011best}
\begin{equation}
\label{PR}
P_i = (1-\alpha)\sum_j P_j \frac{w_{ji}}{s_{j}^{out}} + \frac{\alpha}N +
\frac{1-\alpha}{N}\sum_j P_j\delta(s_j^{out})
\end{equation}
where $P_i$ is the prestige score of team/player/node $i$, $N$ is the number of nodes in the network, $w_{ji}$ is weight of the link from $j$ to $i$ which is equal to the number of wins of $i$ over $j$, $s_j^{out}=\sum_i w_{ji}$ is the out-strength of node $j$ (the number of losses of $j$) and $\alpha$ is the algorithm parameter (often referred to as the teleportation probability).
The last term in \eref{PR} makes the algorithm robust against the nodes with $s_j^{out}=0$ (``dangling nodes'') which would otherwise act as score sinks. In line with~\citep{radicchi2011best} and other PageRank literature, we use $\alpha=0.15$. As it has been widely applied to sports results~\citep{gleich2015pagerank}, $\PageRank$ and its performance is the main focus of this work.

In addition to standard PageRank, we consider here a new method closely based on PageRank which we refer to as bi-directional PageRank ($\BiPageRank$). The bi-directional PageRank score, $S_i$, is defined as
\begin{equation}
S_i = P_i - Q_i
\end{equation}
where $P_i$ is the previously introduced PageRank score (computed on the original directed network) and $Q_i$ is given by
\begin{equation}
Q_i = (1-\alpha)\sum_j Q_j\frac{w_{ij}}{s_{j}^{in}}+\frac{\alpha}{N}+
\frac{1-\alpha }{N}\sum_{j}Q_{j}\delta(s_j^{in})
\end{equation}
In agreement with the previous definition of $w_{ji}$, here $w_{ij}$ is the number of losses of $i$ against $j$ and $s_j^{in}=\sum_i w_{ij}$ is the in-strength of $j$ (the number of wins of $j$). In this way, both the winner of a match is assigned a part of the losing team's score (through $P_i$) as well as the team that loses is assigned a part of the winning team's negative score (through $Q_i$). Bi-directional PageRank is then a simple combination of the two scores. The motivation for this modification is simple: While $P_i$ allows us to award team $i$ ``good'' score based on which teams it won against, $Q_i$ allows us to award team $i$ ``bad'' score based on which teams it lost against. As a practical illustration, take teams $i$ and $j$ that lost all their matches, hence they receive identical PageRank score. If team $i$ lost against good teams (teams that lose rarely) and team $j$ lost against bad teams (teams that lose often), then $S_i>S_j$. The new algorithm thus allows us to distinguish the two teams. Note that separate win and loss scores have been considered also in \citep{park2005network}.

\subsection{Evaluation metrics}
On synthetic data, the resulting ranking of teams produced by a ranking algorithm can be directly compared with their fitness values which are thus used as the ground truth. Denote the computed and ground-truth ranking of team $i$ as $x_i$ and $g_i$, respectively. We use the following distinct metrics to quantify the ranking performance of an algorithm:

\begin{enumerate}
\item The Kendall correlation coefficient, $\tau$, is defined as
\begin{equation}
\label{kendall}
\tau(\boldsymbol{x}, \boldsymbol{g})=\frac{\lvert\{(i, j):\ (x_i - x_j)(g_i - g_j) > 0\}\rvert - \lvert\{(i, j):\ (x_i - x_j)(g_i - g_j)<0\}\rvert}{\tfrac12 n(n-1)}
\end{equation}
where $\boldsymbol{x}$ and $\boldsymbol{g}$ are vectors of the 
In the numerator, the first and second terms correspond to the number of ``concordant'' (their order agrees between the computed and ground-truth ranking) and ``discordant'' (their order differs between the computed and ground-truth ranking) pairs of teams. Kendall's $\tau$ ranges from $+1$ when the rankings are identical to $-1$ when one ranking is the reverse of the other. Note that tied ranking positions in the computed ranking (the ground-truth ranking has no ties by construction) do not contribute to $\tau$; a degenerate ranking that would assign the same rank to all teams would thus achieve $\tau=0$.

\item While Kendall's $\tau$ takes all teams into account, the other two metrics explicitly focus on how well the top teams are ranked. Firstly, the average top-5 ranking is the average computed ranking of the top 5 ground-truth teams. The smaller the value, the better the computed ranking.

\item Secondly, the area under the ROC-curve, commonly referred to as $AUC$ in the statistics literature. We again use the top 5 ground-truth teams as our goal top set; the other teams constitute the ordinary set. To compute $AUC$, we use the probabilistic approach~\citep{lu2011link} where we pick $n$ pairs of teams, one from the top set and the other from the ordinary set. If the top team is higher ranked than the ordinary team $n'$ times and tied $n''$ times, the $AUC$ value can be computed as
$AUC=(n'+n''/2) / n$.
\end{enumerate}

\begin{table}
\resizebox{\textwidth}{!}{
\begin{tabular}{llllrr}
\hline
Sport & Country & League Name & Label & Size & Years\\
\hline
\multirow{4}{*}{Baseball}   & U.S.A      & Major League Baseball -- American League & AL & 14--15 & 1997--2016\\ \cline{2-6}
& U.S.A      & Major League Baseball -- National League & NL & 15--16 & 1997--2016\\ \cline{2-6} 
& Japan      & Nippon Professional Baseball            & NPB           & 12--14                      & 2010--2019\\ \cline{2-6} 
& Mexico     & Ligue Mexicaine de Baseball             & LMB           & 14--19                      & 2007--2019\\ \hline
\multirow{3}{*}{Ice hockey}     & U.S.A      & National Hockey League                  & NHL           & 30--31 & 2000--2019\\
\cline{2-6} 
& Swizerland & Ligue Nationale A                       & LNA           & 12--22 & 2008--2019\\
\cline{2-6}
& Germany    & Deutsche Eishockey Liga                 & DEL           & 14--16 & 2007--2020\\
\hline
\multirow{7}{*}{Soccer}     & Germany    & Deutsche Futball Liga                   & Bundesliga & 18 & 2000--2019\\ \cline{2-6} 
& Italy      & Lega Serie A                            & Serie A       & 20    & 2005--2019\\ \cline{2-6} 
& Spain      & Primera division de Liga                & Liga 1        & 20    & 1998--2017\\ \cline{2-6} 
& England    & England Premier League                  & EPL           & 20    & 1999--2018\\ \cline{2-6} 
& U.S.A      & Major League Soccer                     & MLS           & 10--24 & 2000--2019\\ \cline{2-6} 
& France      & Championnat de France de football Ligue 1                     & Ligue 1           & 20 & 2000--2019\\
\cline{2-6} 
& China      & Chinese Football Association            & CSL           & 12--16 & 2004--2019\\ \hline
\multirow{4}{*}{Basketball} & Italy      & Lega Basket Serie A                     & LBSA       & 15--18                      & 2008--2019\\
\cline{2-6}
& China      & Chinese Basketball Association          & CBA           & 17--20                      & 2007--2017\\
\cline{2-6}
& Spain      & Asociacion de Clubes de Baloncesto      & ACB           & 17--18                      & 2007--2019\\
\cline{2-6}
& U.S.A      & National Basketball Association         & NBA           & 30                         & 2001--2020\\
\hline
\end{tabular}}
\caption{Basic description of the analyzed sports results sets. The league size is the number of competing teams (a~range is provided if the number varies in the considered period).}
\label{tab:datasets}
\end{table}

\subsection{Empirical sports results data}
\label{sec:data}
The analyzed sports data have been obtained from websites \url{https://www.sports-reference.com/} and \url{http://www.win007.com/}. Except for soccer, all games have only two possible outcomes: the home team wins or the away team wins. As we consider an outcome model without draws, all draws (20--25\% of games in one season for all six considered leagues) are ignored. In one season, the participating teams play against each other two or more times. We analyze only results from regular seasons, playoff matches are ignored. See Table~\ref{tab:datasets} for an overview of the analyzed datasets and their basic statistical properties.

\section{Model calibration on real datasets}
To determine realistic values of all $N+2$ model parameters (team fitness values, $H$, and $\delta$), we use maximum likelihood estimation for sets of results from various sports. For a given set of games with outcomes, $\mathcal{G}$ and a single game as $g\in\mathcal{G}$, we denote the home team as $h_g$, the away team as $a_g$, and the game result as $R_g$ where $R_g=1$ means that the home team won game $g$ and $R_g=0$ means that the home team lost. The data likelihood given the model then has the form
\begin{equation}
\label{likelihood}
\mathcal{L}(\mathcal{G}\vert f_1,\dots,f_N,H,\delta) =
\prod_{g\in\mathcal{G}} \Big\{R_g P(h_g, a_g) + (1 - R_g)\big[1 - P(h_g, a_g)\big]\Big\}.
\end{equation}
Likelihood maximization for the real datasets described in Section~\ref{sec:data} reveals a surprising fact that maximum likelihood estimates (MLE) of team fitness values, $\hat f_i$, are close to the fraction of wins, $w_i$, of each respective team in the analyzed dataset. In particular, the difference between the likelihood maximized over all model parameters and the likelihood maximized only over $H$ and $\delta$ (whereas $f_i$ is replaced with $w_i$) is not sufficient to justify the higher number of parameters in the former model ($N+2$ vs. $2$, we used the Akaike information criterion for model selection~\citep{claeskens2008model}). This allows us to write the simplified winning probability of the home team as
\begin{equation}
\label{prob_win_simplified}
P(i,j)=\bigg[1+\ee^{-(\Delta w_{i,j}+H)/\delta}\bigg]^{-1}
\end{equation}
where $\Delta w_{i,j} := w_i - w_j$ is the difference between the win ratios of the home and away teams.

If the home advantage is absent ($H=0$), denoting $\exp[f_i/\delta]:=p_i$ and $\exp[f_j/\delta]:=p_j$ transforms \eref{prob_win} to $P(i,j)=p_i / (p_i + p_j)$ which is precisely the form assumed by the seminal Bradley-Terry model~\citep{bradley1952rank}. We see now that the model formulation presented by \eref{prob_win} is still advantageous as: (1) Unlike the ``winning propensities'' $p_i$, the team fitness values $f_i$ directly correspond to the team win ratios, (2) $H$ introduces home advantage in a scale that can be directly compared with the teams' win ratios and their differences ($H=0.1$ is as important as a $0.1$ difference in win ratios between the teams).

\begin{figure}
\centering
\includegraphics[scale=0.38]{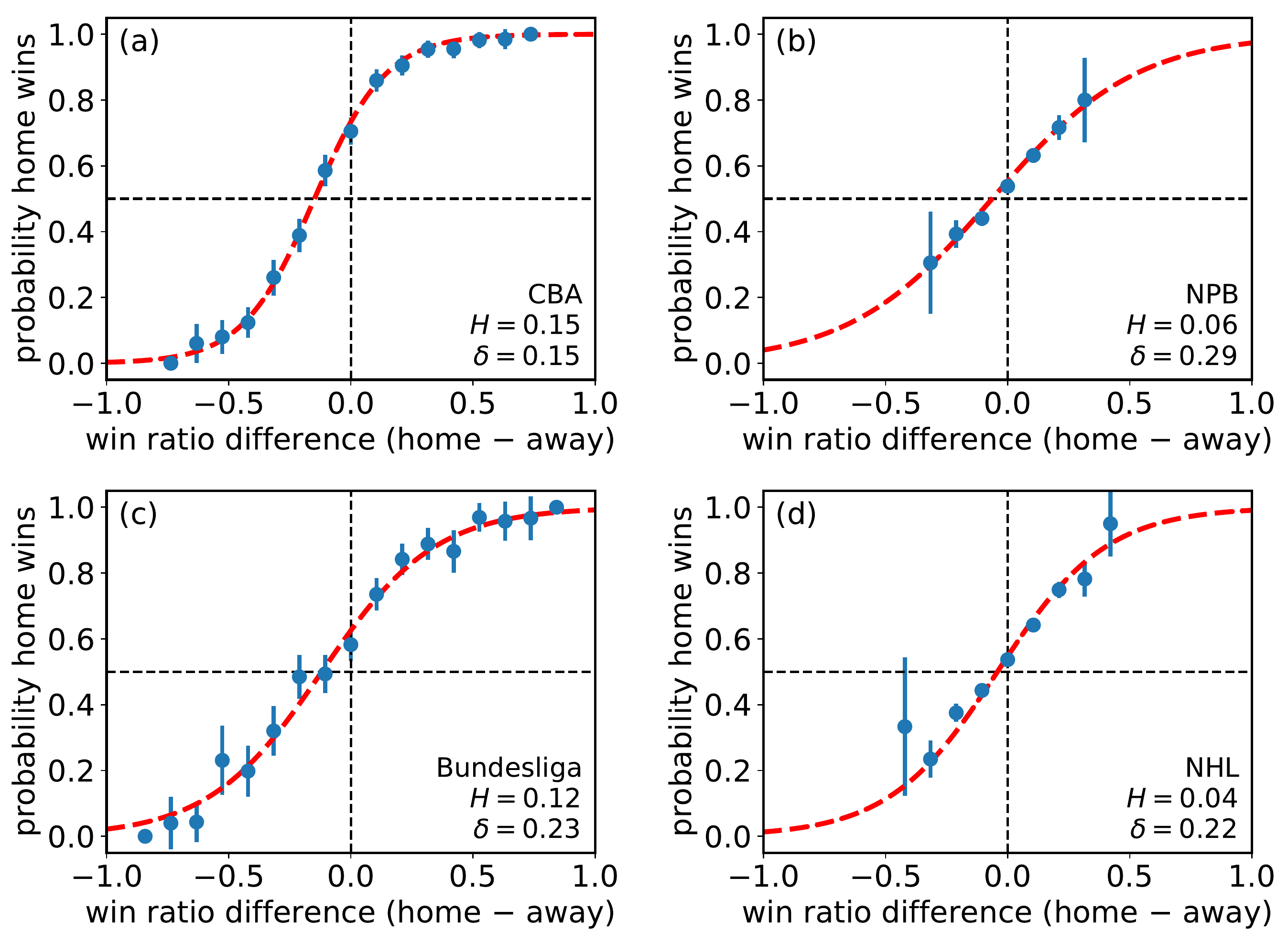}
\caption{Relationship between the win ratio difference and the home team winning probability for four distinct leagues. Symbols show empirical winning probabilities for a given end of season win ratio difference (data points based on less than four games have been omitted), error bars show double of the standard error of the mean, and lines show the model probability of winning given by \eref{prob_win_simplified} for the maximum likelihood estimates of $\delta$ and $H$ (shown in each panel). The horizontal and vertical dashed lines show the zero win ratio difference and the baseline win probability of $1/2$, respectively.}
\label{fig1}
\end{figure}

Using four sample results sets, Figure~\ref{fig1} shows a comparison between \eref{prob_win} with maximum likelihood estimates for $H$ and $\delta$ and the empirical winning probability plotted as a function of the win ratio difference between the competing teams. The good agreement that can be observed in the whole range of win ratio difference confirms that \eref{prob_win} can model the empirical data well. The sigmoid curve's steepness in the figure is in direct relation with the fitness sensitivity parameter $\delta$ (smaller $\delta$ yields higher steepness). Note also that $P(i,j) > 0.5$ when $\Delta w_{i,j} = 0$ which is a direct consequence of a positive home advantage in all four results sets.

\begin{figure}
\centering
\includegraphics[scale=0.38]{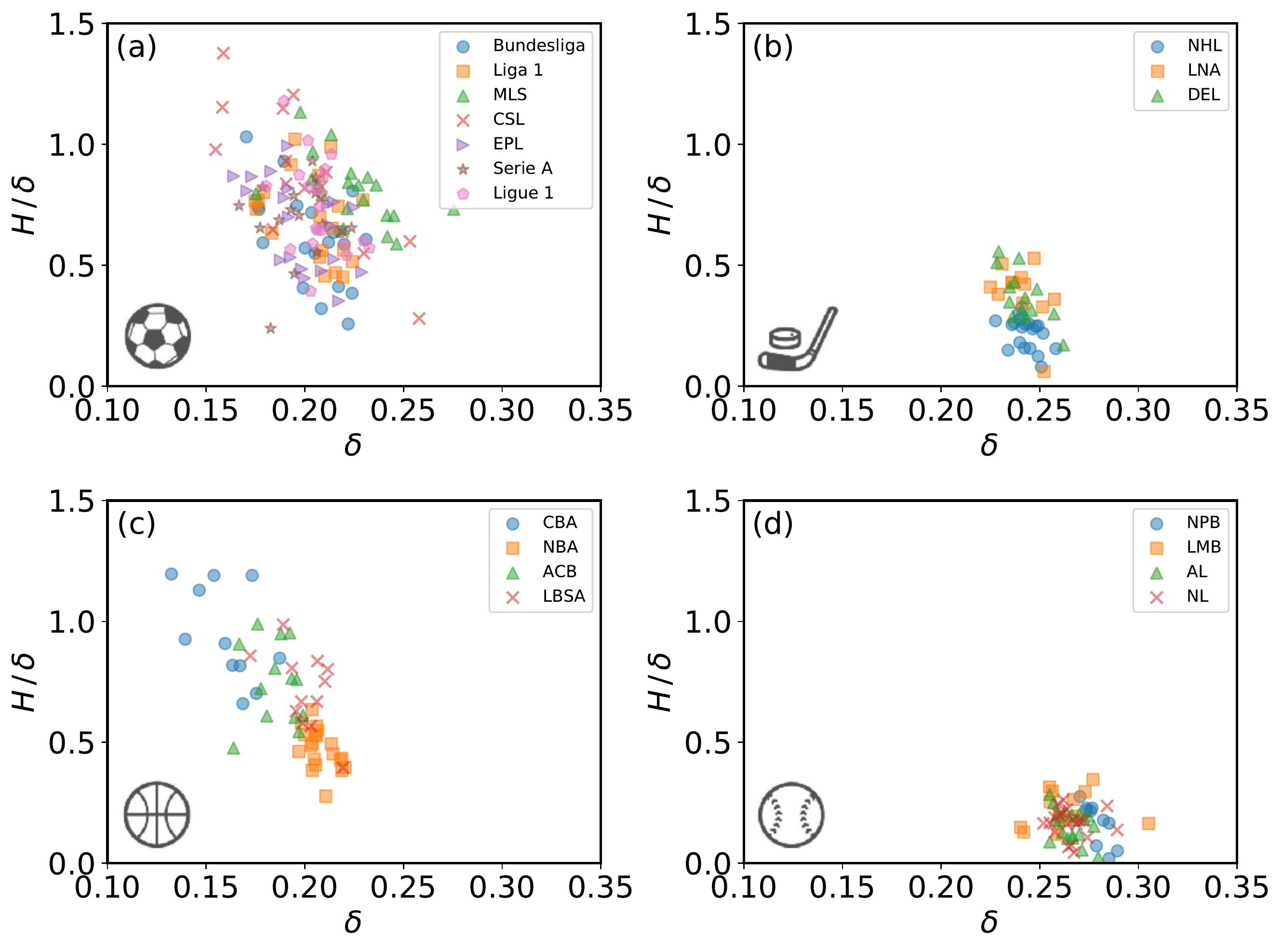}
\caption{Parameter estimates for the sports results sets from Table~\ref{tab:datasets} (each panel shows a different sport). Each symbol represents the estimates of $\delta$ and $H$ in a single season.}
\label{fig2}
\end{figure}

Figure~\ref{fig2} further summarizes the maximum likelihood parameter estimates in all analyzed results sets, divided in panels by the sport kind. We see that different sports differ in their level of randomness as characterized by $\delta$ (baseball and basketball are the most and the least random sport, respectively). Different leagues in the same sport have mostly similar $\delta$ values except for the basketball leagues NBA (U.S.A.) and CBA (China) where CBA is significantly less random than NBA (in fact, CBA is the least random league on average among the analyzed 17 leagues). The home advantage value is distributed between 0 and 0.25, and the home advantage effect of basketball and football is more significant than baseball and hockey. In agreement with \eref{prob_win}, the effective strength of the home advantage is characterized by $H/\delta$ which is shown in Figure~\ref{fig2} on the vertical axes. The values of $H/\delta$ differ significantly between the sports as well as between different leagues in the same sport. CBA is again outstanding by having the highest average effective home advantage. By contrast, baseball leagues have average effective home advantage 5.4-times smaller than CBA.

\begin{figure}
\centering
\includegraphics[scale=0.38]{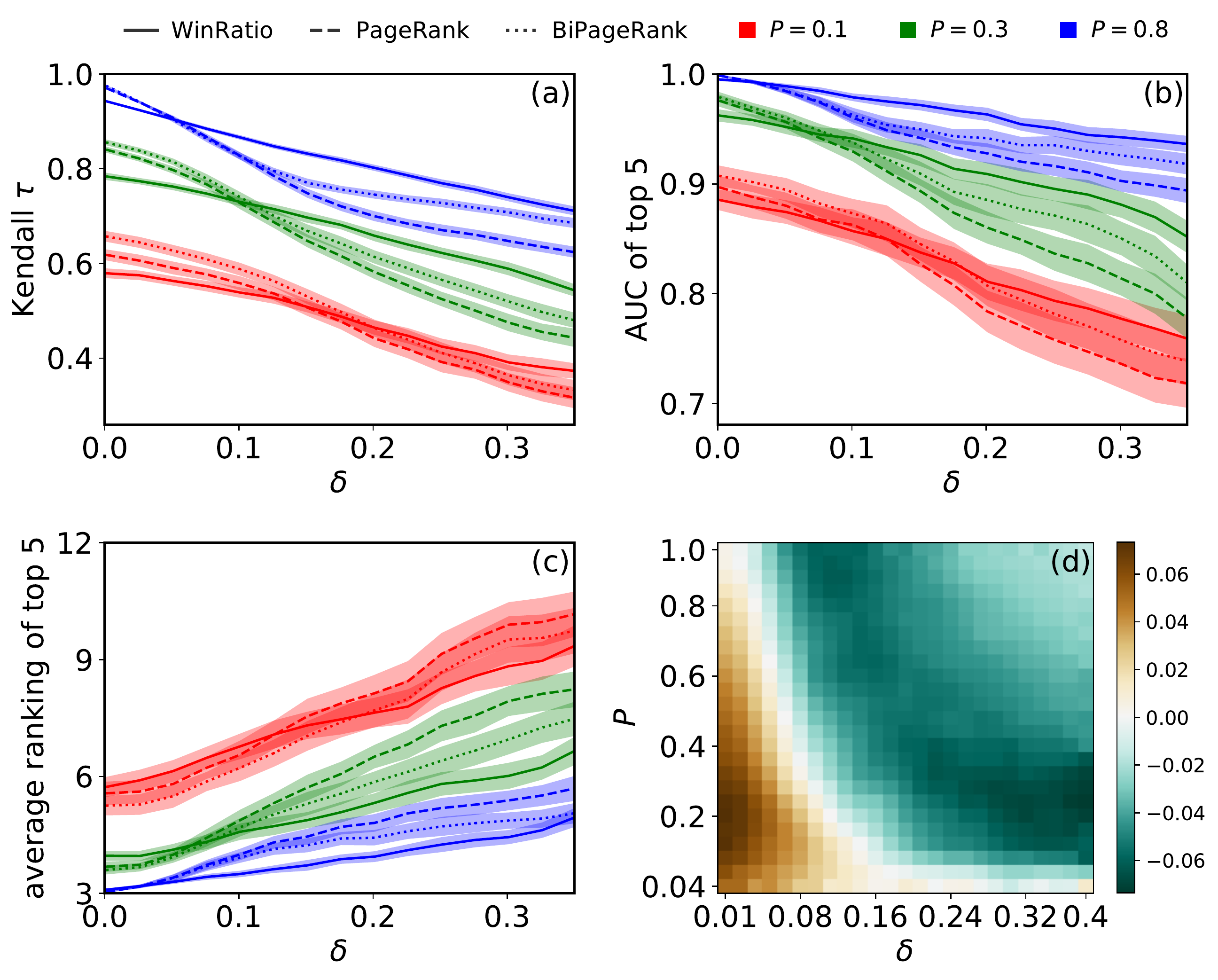}
\caption{Performance of the studied ranking algorithms on synthetic data without home advantage. Panels (a)--(c) use three different evaluation metrics and plot the results as a function of fitness sensitivity, $\delta$, for various fractions of played matches, $P$. The lines represent mean results and the shaded regions indicate double of the standard error of the mean, all determined from 100 independently created synthetic datasets.
Panel (d) shows the Kendall $\tau$ difference between Bi-directional PageRank and the win ratio as a function of both $\delta$ and $P$.}
\label{fig3}
\end{figure}

\section{Results on synthetic datasets}
Our goal now is to assess the performances of different ranking algorithms on synthetic sports results generated by the above-described algorithm. In simulations, we assume that there are 30 teams; the results are robust with respect to the number of teams. We begin by studying the case of no home advantage ($H=0$) and explore a range of $\delta$ values which corresponds to the empirical values in Figure~\ref{fig2}.

Figure~\ref{fig3} shows the results of numeric simulations comparing the three considered ranking algorithms as a function of fitness sensitivity, $\delta$, and the fraction of matches played, $P$. Panels a--c show that the comparison results are remarkably similar for the three evaluation metrics (Kendall's tau, $AUC$ and average ranking). In particular, the results show that: (1) As $P$ grows, the ranking performance improves as expected. (2) PageRank outperforms the win ratio only when $\delta$ is sufficiently small and the range of PageRank's superiority shrinks as $P$ grows. The threshold $\delta$ below which PageRank outperforms the win ratio is considerably stable with respect to the number of teams, $N$ (results not shown). (3) The newly-proposed Bi-directional PageRank is always an improvement (or a tie) over standard PageRank. (4) Figure~\ref{fig2} shows that a vast majority of the analyzed datasets have $\delta>0.15$ which together with Figure~\ref{fig3} implies that the use PageRank brings no improvement in sport tournament rankings. Even more, PageRank is significantly inferior for sports with high randomness (high $\delta$) later in a season. Finally, the heatmap in Figure~\ref{fig3}(d) provides a comparison between bi-directional PageRank and the win ratio for a broad range of the key parameters $\delta$ and $P$. We can see here well that a tie between the two ranking algorithms occurs at $\delta$ which progressively decreases as $P$ grows.

Based on Figure~\ref{fig3}, we can conclude that PageRank and bi-directional PageRank are both more sensitive than the win ratio to increasing randomness of outcomes (represented by increasing $\delta$). This increased sensitivity can be explained by the algorithms' network nature: While a ``surprise'' outcome of a single match has only a local impact on the win ratio (only the two competing teams are affected), PageRank propagates its scores further over the whole network of teams. When $\delta$ is sufficiently large, the surprising outcomes are numerous and their network propagation and accumulation are ultimately detrimental to the ranking ability of PageRank and Bi-directional PageRank.

\begin{figure}  
\centering	
\includegraphics[scale=0.38]{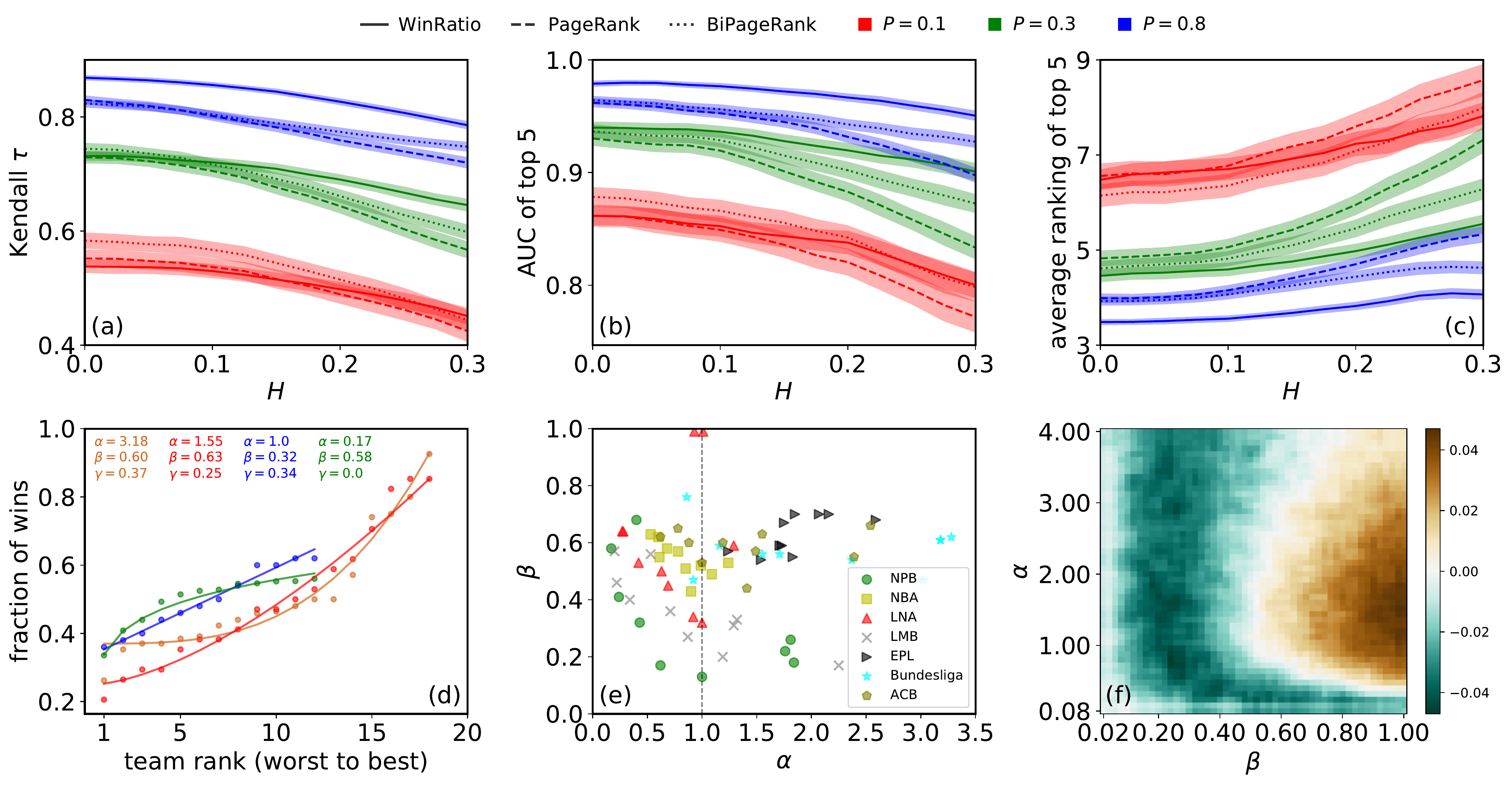}
\caption{Performance of the studied ranking algorithms on synthetic data with home advantage (top row) and with a non-uniform distribution of team fitness (bottom row).
Panels (a)--(c) show the ranking performance vs. the home advantage parameter, $H$, for fixed $\delta=0.1$ (all other parameters as in Figure~\ref{fig3}). (d) The relation between the fraction of wins and the team rank (from the worst to the best) for four chosen result sets (NHL, 2011; ACB, 2011; LMB, 2012; Bundesliga, 2010). Panel (e) Fits of \eref{nonlinear} for all seasons of seven different leagues. (f) The difference $\tau(\BiPageRank)-\tau(\WinRatio)$ for synthetic data with various values of $\alpha$ and $\beta$ ($\delta=0.1$, $H=0$, and $P=0.1$, results are averaged over 100 realizations).}
\label{fig4}
\end{figure}

The top row of Figure~\ref{fig4} evaluates the ranking performance of algorithms when the home advantage parameter, $H$, is positive. We see that as $H$ grows, the performance of all three ranking algorithms deteriorates. At the same time, the win ratio is more robust to increasing $H$ than the other two ranking methods, which is in line with its higher robustness to increasing $\delta$. In particular, the number of unexpected results (a weaker team wins against a stronger team) increases as $H$ grows and these unexpected results negatively affect the ranking results of PageRank and Bi-directional PageRank. The home advantage thus further reduces the limited range of applicability of PageRank (the range in which PageRank outperforms the win ratio).

In synthetic data so far, we assumed the team fitness values to be uniformly assumed. That this is not the case in real data can be easily illustrated as we have already shown that the win ratio is a good approximation for team fitness. Figure~\ref{fig4}d shows the win ratio in four different datasets and shows distinct non-linearity for two of them. This motivates us to consider a non-linear assignment of fitness in the form
\begin{equation}
\label{nonlinear}
f_i = \beta\left(\frac{i-0.5}{N}\right)^{\alpha} + \gamma
\end{equation}
which fits well most of the considered datasets (see least-squares fits in Figure~\ref{fig4}d). A power-law fitness distribution has been suggested also before in~\citep{da2013hidden}. In~\eref{nonlinear}, $\alpha$ controls the heterogeneity of the fitness distribution and $\beta$ determines the difference between the worst and the best team. Once $\alpha$ and $\beta$ are chosen, $\gamma$ is fixed by the relation $\sum_i f_i /N = 1/2$ (the average win ratio must be one half as someone's win is always someone else's loss).\footnote{Also, the fitness difference $f_i-f_j$ that influences the match outcome in \eref{prob_win} is directly determined by $\alpha$ and $\beta$ as the absolute term $\gamma$ cancels out.}

Figure~\ref{fig4}e further shows the fitted values $\alpha$ and $\beta$ for seven representative leagues and helps us identify $\alpha\in(0, 3.5)$ and $\beta\in(0, 1)$ as the relevant ranges for these two parameters. Figure~\ref{fig4}f shows the difference between the win ratio and bi-directional PageRank for synthetic data generated with $\alpha$ and $\beta$ in the identified range. We choose here by purpose parameters that favor bi-directional PageRank: small randomness ($\delta=0.1$), no home advantage ($H=0$) and few games played ($P=0.1$). In agreement with the results presented in Figure~\ref{fig3}, $\BiPageRank$ outperforms $\WinRatio$ when $\alpha=1$ and $\beta=1$ (fitness values are then uniformly distributed in the range $[0,1]$). We see now that this is essentially the ideal setup for $\BiPageRank$ as its advantage decreases when $\alpha$ substantially differs from $1$ as well as when $\beta$ is lower than $1$. This is because the average fitness difference between the teams then decreases which, in agreement with \eref{prob_win}, increases the probability of unexpected outcomes. Thus-introduced randomness is detrimental to the performance of PageRank and bi-directional PageRank which is well visible in Figure~\ref{fig4}e. When $\delta$, $H$, or $P$ increase, the behavior is similar, only the region where $\BiPageRank$ outperforms $\WinRatio$ shrinks.

\begin{figure}
\centering
\includegraphics[scale=0.38]{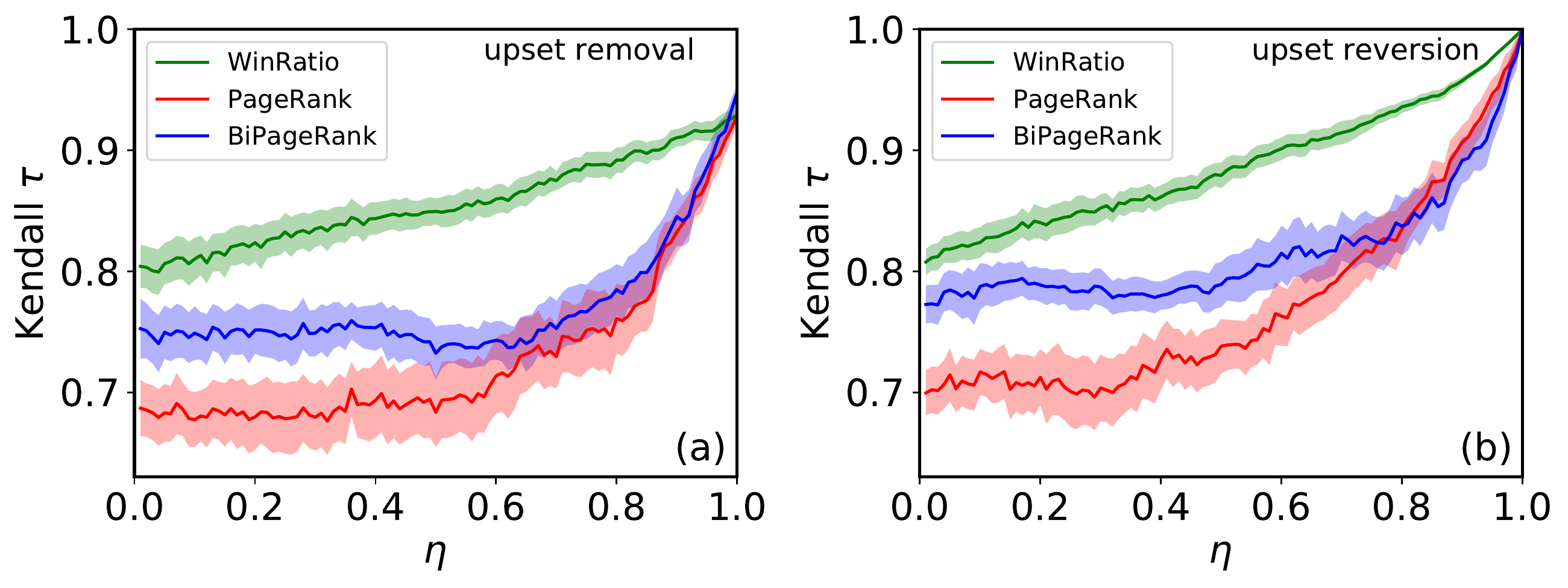}
\caption{Ranking performance of the evaluated algorithms when fraction $\eta$ of unexpected outcomes ($i$ wins over $j$ when $f_i<f_j$) are: (a) removed, (b) reverted. Simulation parameters: $N=30$, $\delta=0.25$, $H=0.08$, $P=1.0$, $\alpha=1.5$, $\beta=0.5$, results are averaged over 100 independent realizations of the model.}
\label{fig5}
\end{figure}

In summary, we identified the sensitivity of PageRank and bi-directional PageRank to unexpected results as the main factor limiting their performance. Unexpected results are due to intrinsic randomness of sport, home advantage, and similarity of team fitness values (in reality, many other factors contribute---weather, immediate form of individual players, injuries, and others). As our initial empirical analysis shows, all these factors are common to sports results data. If substantial randomness of results is inevitable, one can ask if we can at least suppress the unexpected results to help $\PageRank$/$\BiPageRank$ perform better and possibly outperform the win ratio. To explore the feasibility of this idea, we benefit from the use of synthetic data where team fitness values are known. We can thus identify the unexpected results (wins of weaker teams against stronger teams) and either remove them from the dataset (see Figure~\ref{fig5}a) or reverse them (see Figure~\ref{fig5}b). In Figure~\ref{fig5}, we remove or correct a gradually increasing fraction of unexpected outcomes ($\eta=1$ means that all unexpected outcomes have been treated) which naturally benefits all three evaluated algorithms. However, PageRank and bi-directional PageRank require large $\eta$ for their performance to improve substantially whereas the win ratio improves uniformly in the whole range of $\eta$. As a result, there is no $\eta$ for which PageRank or bi-directional PageRank perform better than the win ratio. In real data where team fitness values are not directly known, one would first have to identify the unexpected results, which would further lower the efficiency of this approach. We can thus conclude that the removal or correction of unexpected results cannot help PageRank and bi-directional PageRank outperform the win ratio.

\section{Results on real datasets}
After comparing the ranking performance on synthetic datasets in the previous section, we now present a similar comparison on real datasets. Since team fitness values are not known in real data, we use the ranking of all teams at the end of the season as the ground truth against rankings produced by respective ranking algorithms in earlier parts of the season. This choice is motivated by the earlier observation that the team win ratio is a good proxy for team fitness [see the discussion before \eref{prob_win_simplified}]. Denoting the number of games in season $s$ as $N_S$, we then use first $PN_S$ games as input for algorithm $A$ and quantify the algorithm's performance using Kendall's $\tau$ between the computed ranking and the end of season number of wins, thus obtaining $\tau_S(P,A)$. This is then averaged over seasons to produce $\tau(P,A)$. In this way, we compare the performance of the win ratio with that of bi-directional PageRank by evaluating $\tau(P,\BiPageRank)-\tau(P,\WinRatio)$.

\begin{figure}
\centering
\includegraphics[scale=0.55]{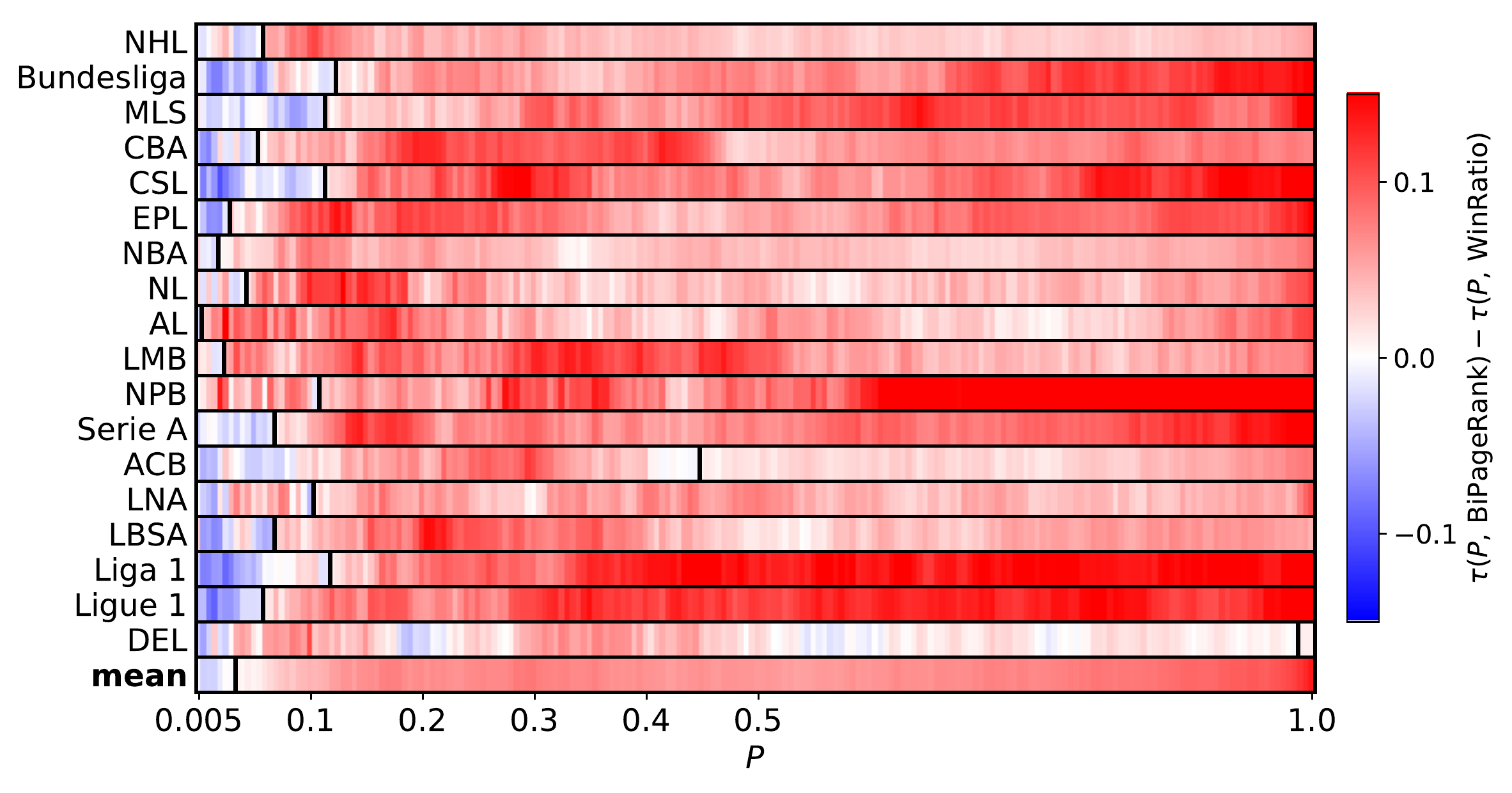}
\caption{The performance difference $\tau(P,\BiPageRank)-\tau(P,\WinRatio)$ in real sports data where the final win ratio in each season is used as the ground truth (the results are averaged over the last 10 available seasons). Individual rows represent different leagues from the beginning ($P=0$, left) until the end ($P=1$, right). Vertical markers highlight the highest $P$ value where $\BiPageRank$ outperforms $\WinRatio$.}
\label{fig6}
\end{figure}

The results are shown in Figure~\ref{fig6} where each horizontal bar represents one league with left and right sides representing the start and the end of the season, respectively, and the computed differences between $\BiPageRank$ and $\WinRatio$ are color-coded. We see that despite using the final fraction of wins in a season as the ground truth (which obviously favors $\WinRatio$), $\PageRank$ is still able to outperform $\WinRatio$ when $P$ is small. In a direct parallel with previously presented results on synthetic data, the win ratio outperforms bi-directional PageRank almost always except for the very beginning of the season (small $P)$. To highlight the transition between the early part of the league when BiPageRank is best and the later part when the win ratio is best, we mark the largest $P$ at which $\tau(P,\BiPageRank)-\tau(P,\WinRatio)>0$ with a vertical line for each league. These threshold $P$ values are around 0.1 or lower except for two leagues (ACB and DEL) for which narrow ranges with weakly positive $\tau(P,\BiPageRank)-\tau(P,\WinRatio)$ appear also for large $P$. The overall behavior is best visible in the last row where the difference between $\BiPageRank$ and $\WinRatio$ is averaged over all considered leagues. The threshold $P$ value here is $0.035$ and $\WinRatio$ never outperforms $\BiPageRank$ by more than $0.03$ (as measured by Kendall's $\tau$). This confirms in a model-free way that PageRank and bi-directional PageRank are beneficial for sports results data only when the information is scarce ($P$ is low). When sufficiently many teams have already played against each other, the win ratio generally ranks the teams better.

\section{Conclusions}
In this paper, we have focused on sports results data from regular leagues where a fixed number of teams play against each other. Results of the games can be represented as a directed network where a directed link from $i$ to $j$ is drawn when team $j$ has won over $i$. We have evaluated the ranking performance of three distinct methods to rank the competing teams: their win ratio, their PageRank score, and their newly proposed bi-directional PageRank score. Bi-directional PageRank combines two different scores: one positive which accumulates mainly through winning over good opponents (as in PageRank), the other negative which accumulates mainly through losing against bad opponents. We have calibrated a model for synthetic sports results, a variant of the classical Bradley–Terry model~\citep{bradley1952rank}. The model uses only two parameters, home advantage $H$ and sport randomness $\delta$, yet it produces excellent agreement with empirical sports results.

The ranking algorithms have been first evaluated on synthetic data. The main finding is that PageRank only outperforms the win ratio when a small fraction of all games have been played \emph{and} randomness of results are sufficiently small. In particular, PageRank yields for the levels of randomness found in real sports data (we considered baseball, ice hockey, soccer and basketball). Note that while~\citep{ghoshal2011ranking} reports that incompleteness of the network is harmful to PageRank's performance, which is natural, we find that incompleteness of the network
is actually favorable when PageRank's performance is judged relative to the win ratio benchmark.

The newly proposed $\BiPageRank$ outperforms $\PageRank$ for all parameter settings, yet it only outperforms $\WinRatio$ only for sports with the lowest randomness when a small fraction of all games (10\%) have been played. Both $\PageRank$ and $\BiPageRank$ further suffer when other sources of randomness---home advantage and non-uniform distribution of team abilities---are considered. The sensitivity of $\PageRank$, and closely related $\BiPageRank$, to changes in the data has been already discussed in, for example, \citep{chartier2011sensitivity}. We demonstrate here, for the first time, that this sensitivity combined with the natural randomness of sport renders PageRank of little use on results from a sport tournament. By contrast, the ranking of teams by their win ratio turns out to be comparatively robust to various sources of randomness in results.

To keep the model for synthetic sports results simple, we neglected further factors that can be addressed in future research. The assumption of fixed team fitness can be relaxed to allow for modeling a variable sport level or temporary adverse effects of injuries, for example. The simple tournament setup can be generalized to irregular games between the teams as is the case for national teams in soccer or players in tennis, for example. In tennis, in particular, each player has recently played only a small fraction of other players. Using the terminology of our model, the effective $P$ is small, which suggests that PageRank might have some merit for tennis data.

Besides providing specific results on the use of PageRank on sports results data, our work highlights the need to carefully assess the actual performance and limitations of network metrics. This need is exacerbated by the complexity of systems that produce the data, which makes it difficult to judge \emph{ex-ante} if an algorithm is a good match for the data. In citation data, for example, PageRank has been frequently used yet \citep{mariani2015ranking} shows that the natural growth of the citation network makes PageRank scores difficult to interpret. If a ground truth set is available, a comparative assessment on real data is possible. This can be made more robust by considering multiple real datasets and multiple ground truth sets as done recently in \citep{xu2020unbiased} to compare ranking metrics for citation data. If a ground truth set is not available but a credible model for a given system exists, an assessment using synthetic data (as we have used here) is a practical alternative. Using a network metric without understanding its scope and limitations directly induces the risk of obtaining unreliable or inferior results.

\section*{Acknowledgement}
This work was supported by the Swiss National Science Foundation (grant No.~182498). MM and AZ acknowledge support from the National Natural Science Foundation of China (grant Nos.~11850410444 and 71731002, respectively).

\bibliography{cas-refs}
\end{document}